\documentstyle[prb,aps,epsf,eqsecnum,prbbib]{revtex}

\begin{document}
\draft
\twocolumn[\hsize\textwidth\columnwidth\hsize\csname  
@twocolumnfalse\endcsname
\title{From superconducting fluctuations to the bosonic limit in 
the response functions above the critical temperature}
\author{G.C. Strinati and P. Pieri}
\address{
Dipartimento di Matematica e Fisica, Sezione INFM\\ 
Universit\`{a} di Camerino, I-62032 Camerino, Italy}

\date{\today}
\maketitle
\hspace*{-0.25ex}

\begin{abstract}
The density, current, and spin response functions are investigated 
\emph{above\/} the critical temperature for a system of three-dimensional 
fermions interacting via an attractive short-range potential, as the strength
of this potential is varied from weak to strong coupling.
In the strong-coupling (bosonic) limit, we identify the 
dominant diagrammatic contributions for a ``dilute'' system of composite 
bosons which form as bound-fermion pairs, by giving appropriate prescriptions
for mapping bosonic onto fermionic diagrams. We then extrapolate these
contributions to the weak-coupling limit and compare them with the 
ordinary (Aslamazov-Larkin, Maki-Thompson, and density-of-states) terms 
occurring in the theory of superconducting fluctuations for a clean system 
above the critical temperature.
Specifically, we show that in the strong-coupling limit, at the zeroth order 
in the diluteness parameter for the composite
bosons, the Aslamazov-Larkin term represents formally the dominant 
contribution to the density and current response functions, while the 
Maki-Thompson and density-of-states terms are strongly suppressed.
Corrections to the Aslamazov-Larkin term are further identified via the above 
mapping prescriptions at the next order in the diluteness parameter for the 
composite bosons, where the residual mutual interaction appears explicitly.
The spin response function is also examined, and it is found to be 
correctly suppressed in the strong-coupling limit only when appropriate sets 
of diagrams are considered simultaneously, thus providing a criterion for 
grouping diagrammatic contributions to the response functions.
\end{abstract}
\hspace*{-0.25ex}
]
\narrowtext
\section{Introduction}

Response functions constitute an essential tool for connecting experimentally 
measurable quantities with the theoretical description of a condensed-matter 
system.
Specifically, knowledge of density, current, spin, and heat response functions
allows one to test the relevance of the degrees of freedom which are selected 
for an approximate description of a complex system.
In particular, for superconductors the current response function plays a 
special role since the Meissner effect can be demonstrated by examining its 
behavior.\cite{Schrieffer}

Within the standard BCS (weak-coupling) theory, the transverse current 
response function below the critical temperature is represented as a 
particle-hole bubble in terms of normal and anomalous 
single-particle propagators.\cite{Schrieffer,FW} 
Above the critical temperature, the noninteracting Fermi gas expression 
is correspondingly obtained, with no sign of superconductivity
being evidenced when approaching the transition from above.

In the weak-coupling limit, precursor effects of superconductivity 
\emph{above\/} the critical temperature have been considered, by 
introducing \emph{pairing fluctuations\/} in the Fermi gas due to the same 
attractive interaction which is responsible for the formation of the superconducting state below 
the critical temperature.
In this way, the so-called Aslamazov-Larkin (AL),\cite{AL} Maki-Thompson (MT),\cite{MT} 
as well as the density-of-state (DOS) contributions have been evaluated and tested 
against experimental data, for superconducting samples of reduced dimensionality 
\cite{Tinkham-75} and for strongly anisotropic cuprate superconductors in the overdoped 
region.\cite {Varlamov}

No corresponding analysis has, however, been performed in the strong-coupling 
limit, where composite bosons form due to the strong fermionic attraction.
Purpose of this paper is to provide this analysis, by setting up a formal 
classification of the diagrammatic structure for the response functions that 
holds specifically in the strong-coupling limit. In this way, one is able to
cover the whole interaction range from weak to strong coupling, by merging 
the two alternative approaches (holding separately in the weak- and 
strong-coupling regimes) through the intermediate-coupling region. This 
merging appears altogether nontrivial, in that two different small 
parameters (namely, the Ginzburg and gas diluteness parameters) control the 
theory in weak and strong coupling.

The intermediate-coupling region might be specifically relevant for 
cuprate superconductors, for which the pairing is likely to be 
in an \emph{intermediate\/} regime between overlapping Cooper pairs and 
non-overlapping composite bosons.
In fact, the small value of the (superconducting) coherence length and the 
presence of a pseudogap above the critical temperature in the underdoped 
region \cite{Ding,Loeser} have suggested a \emph{crossover} scenario, 
from a weak-coupling regime with Cooper pairs forming and condensing at the 
critical temperature within a BCS description, toward a strong-coupling 
regime whereby preformed (composite) bosons exist above the 
superconducting critical temperature and Bose-Einstein condense below 
it.\cite{Leggett,NSR,Randeria,Haussmann,PS,Zwerger,Levin} 

In this context, it appears relevant to study how the response functions evolve
toward the strong-coupling limit, by specifically examining how the response 
of the original Fermi system can be interpreted in terms of the response of 
an effective Bose system.
This evolution of the response functions rests on the property that the 
fluctuation propagator, which constitutes the building block of fluctuation 
theory in the weak-coupling 
limit \emph{above\/} the superconducting critical temperature,
acquires the form of the propagator for composite bosons in the 
strong-coupling limit.

The dominant diagrammatic contributions to the response functions in the 
strong-coupling (bosonic) limit will be selected by relying on the 
diluteness condition of the system (which is automatically satisfied in 
the strong-coupling limit~\cite{Haussmann,PS}), in a similar fashion to what 
was done in Ref.~\onlinecite{Pi-S-98} for the selection of the fermionic 
self-energy. 
In that reference, the diluteness condition was exploited to determine the 
bosonic propagator entering the fermionic self-energy, where the bosonic 
propagator couples with a fermionic propagator. In this paper, we 
apply the diluteness
condition to the physical response functions, for which a description
in terms of bosons will naturally emerge in the strong-coupling limit.

Although the above procedure is \emph{a priori\/} complementary to the 
selection of fluctuation diagrams in the weak-coupling limit, it yet results
into the \emph{same\/} diagrams for the current (and density) response 
functions as far as the dominant contribution (over and above the free 
fermion contribution) is concerned.
Specifically, the AL diagram turns out to yield the dominant contribution to 
the current (and density) response functions {\em both\/} in the weak-coupling 
limit (where it represents the main fluctuation effects close to the critical
temperature) and in the strong-coupling limit (where it corresponds to a 
free-boson response). Corrections to the AL diagram will also be identified in 
the strong-coupling regime at the next-to-leading order in the diluteness 
parameter, thus including interaction effects between composite bosons.
Although the calculation of these corrections  
to the response functions remains still to be implemented numerically, 
their possible relevance to recent results reported in 
Ref.\onlinecite{Perali} will be considered below.

Besides providing a detailed analysis of the current (and density) response 
functions, we will also examine the spin response function.
We shall verify that the diagrams selected in the strong-coupling limit for 
the current (and density) response functions, give an identically 
vanishing contribution to the spin response function, since they correspond to
spinless (composite) bosons. For obtaining a non vanishing contribution in 
the interesting intermediate-coupling region, therefore, the diagram for the
spin response function have to be selected in the weak-coupling region, 
paying, however, attention that their contribution vanishes in the 
strong-coupling limit. To this end, it will be shown how certain diagrammatic
contributions have to be included simultaneously in an appropriate way.

We shall specifically consider a Fermi system with an attractive 
(point-contact) interaction in a three-dimensional continuum and 
\emph{above\/} the superconducting critical temperature.
No lattice or impurities effects will be taken into account.
Consideration of the broken-symmetry case below the critical temperature 
is postponed to future work.

The plan of the paper is as follows.
Section II discusses the current response function at the leading and 
next-to-leading order in the diluteness parameter for composite bosons.
Section III considers the density and spin response functions.
Section IV gives our conclusions.
\section{Current response function}
In this Section, we identify the dominant diagrammatic contributions to the 
current response function for a system of fermions with an attractive 
interparticle interaction in the strong-coupling limit.
The leading contribution turns out to coincide formally with the AL diagram,
occurring in the standard theory of superconducting fluctuations above the 
critical temperature. Next-to-leading diagrams in the bosonic diluteness 
parameter are also considered, to include the effects of the residual 
interaction between the composite bosons in the strong-coupling limit.
Additional diagrams (such as the MT and DOS contributions), which are usually 
considered in superconducting fluctuation theory, are further shown to be 
irrelevant in the strong-coupling limit.
 
The systematic procedure for selecting the contributions to the response 
functions which 
are dominant \emph{in the strong-coupling limit\/} rests on certain 
integrals (that contain products of fermionic single-particle Green's 
functions) acquiring
a particularly simple form in the strong-coupling limit, and on the 
standard classification of bosonic diagrams in the dilute limit.\cite{Popov}
In this way, the contributions to the response functions are organized in powers of the 
diluteness parameter as well as of the (inverse of the) fermionic chemical potential.
The property of the fermionic chemical potential of being 
the largest energy scale in the strong-coupling limit, in fact, considerably 
simplifies dealing with this limit.

The identification of the dominant diagrammatic contributions to the response
functions for a system of fermions in the strong-coupling limit could proceed 
in two ways. Either, one may consider all possible {\em fermionic} diagrams and
estimate their relative contributions by relying on the simplifying features
mentioned above; or, one may start directly from the {\em bosonic} diagrams for
a system of true bosons and construct the corresponding diagrams for 
composite
bosons, where remnants of the underlying fermionic degrees of freedom 
explicitly appear. We shall develop below the latter approach, which has 
by construction the advantage of identifying the important
contributions to the response functions in the strong-coupling limit.

Detailed knowledge of the fermionic attractive interaction 
is not required for studying the evolution from weak 
to strong coupling.
One may accordingly consider a ``contact'' potential
$v_{0} \, \delta ({\mathbf r})$, where $v_{0}$ is a negative constant.
A suitable regularization is required in this case to remove divergences 
in the diagrammatic structure. 
In three dimensions, it is common practice to introduce the fermionic 
\emph{scattering length\/} $a_{F}$ defined via (we set $\hbar$ and 
Boltzmann's constant equal to unity throughout)

\begin{equation}
\frac{m}{4 \pi a_{F}} \, = \, \frac{1}{v_{0}} \, + \, \int \! 
\frac{d{\mathbf k}}
{(2 \pi)^{3}} \, \frac{m}{{\mathbf k}^{2}}  \,\, ,               
\label{ferm-scatt-ampl}
\end{equation}

\noindent
where ${\mathbf k}$ is a wave vector and $m$ the fermionic mass.
The ultraviolet divergence on the right-hand side of 
Eq.~(\ref{ferm-scatt-ampl}) is compensated by letting 
$v_{0} \, \rightarrow \, 0^{-}$ in a suitable way, keeping $a_{F}$ finite.
This is achieved by introducing an ultraviolet cutoff $k_{0}$ and 
choosing $v_{0}$ such that \cite{Pi-S-98}

\begin{equation}
v_{0} \, = \, - \, \frac{2 \pi^{2}}{m k_{0}} \, - \, 
                   \frac{\pi^{3}}{m a_{F} k_{0}^{2}} \,\, ,               \label{vo}
\end{equation}

\noindent
with $k_{0} \rightarrow \infty$ eventually.
The evolution from weak to strong coupling can thus be tuned by varying
the scattering length $a_{F}$, which is negative in the weak-coupling regime (where a 
bound-state has not yet appeared in the associated two-body problem) and positive in 
the strong-coupling (bosonic) regime (where $a_{F}$ coincides with the bound-state radius).

It was discussed in Ref.~\onlinecite{Pi-S-98} that the explicit form 
(\ref{vo}) for $v_{0}$ considerably simplifies the 
structure of the associated many-body perturbation theory.
It was shown there that the effects of the interaction survive \emph{only\/} in the 
particle-particle ladder depicted in Fig.~1(a), while connections among different 
ladders (as well as other links required to form the fermionic self-energy)
are provided by the bare single-particle fermionic Green's functions.
In the context of the correlation functions considered in the present paper, 
a current (or density or spin) vertex made by fermionic 
single-particle Green's functions (cf.~Fig.~1(b)) is further
required to connect the external (electromagnetic) coupling with the structure
of the fermionic two-particle Green's function.

The general expression of the particle-particle ladder of Fig.~1(a) for any 
temperature and value of $a_{F}$ reads:\cite{Pi-S-98}

\noindent
\begin{eqnarray}
& &\Gamma_0(q) \, = \, - \, \left\{ \frac{m}{4 \pi a_{F}} \, +  \,
\int \! \frac{d{\mathbf k}}{(2 \pi)^{3}}\right. \nonumber\\  
& &\times\left.\left[ \phantom{\frac{1}{1}} \frac{\tanh(\beta \xi({\mathbf k})/2) 
+\tanh(\beta \xi({\mathbf k-q})/2)}{2(\xi({\mathbf k})+\xi({\mathbf k-q})-i\omega_{\nu})} 
\, - \, \frac{m}{{\mathbf k}^{2}} \right] \right\}^{-1} 
           \label{general-pp-sc}
\end{eqnarray}

\noindent
with the four-vector notation $q \equiv ({\mathbf q},\omega_{\nu})$, where 
${\mathbf q}$ is a wave vector, $\omega_{\nu} = 2 \nu \pi \beta^{-1}$ ($\nu$ 
integer) a bosonic Matsubara frequency, $\beta = 1/T$ the inverse temperature,
and $\xi({\mathbf k}) = {\mathbf k}^{2}/(2m) - \mu$ ($\mu$ being the fermionic
chemical potential).
This expression acquires a particularly simple form in the strong- and 
weak-coupling limits.

\begin{figure}
\centering
\narrowtext 
\epsfxsize=3.1in
\epsfbox{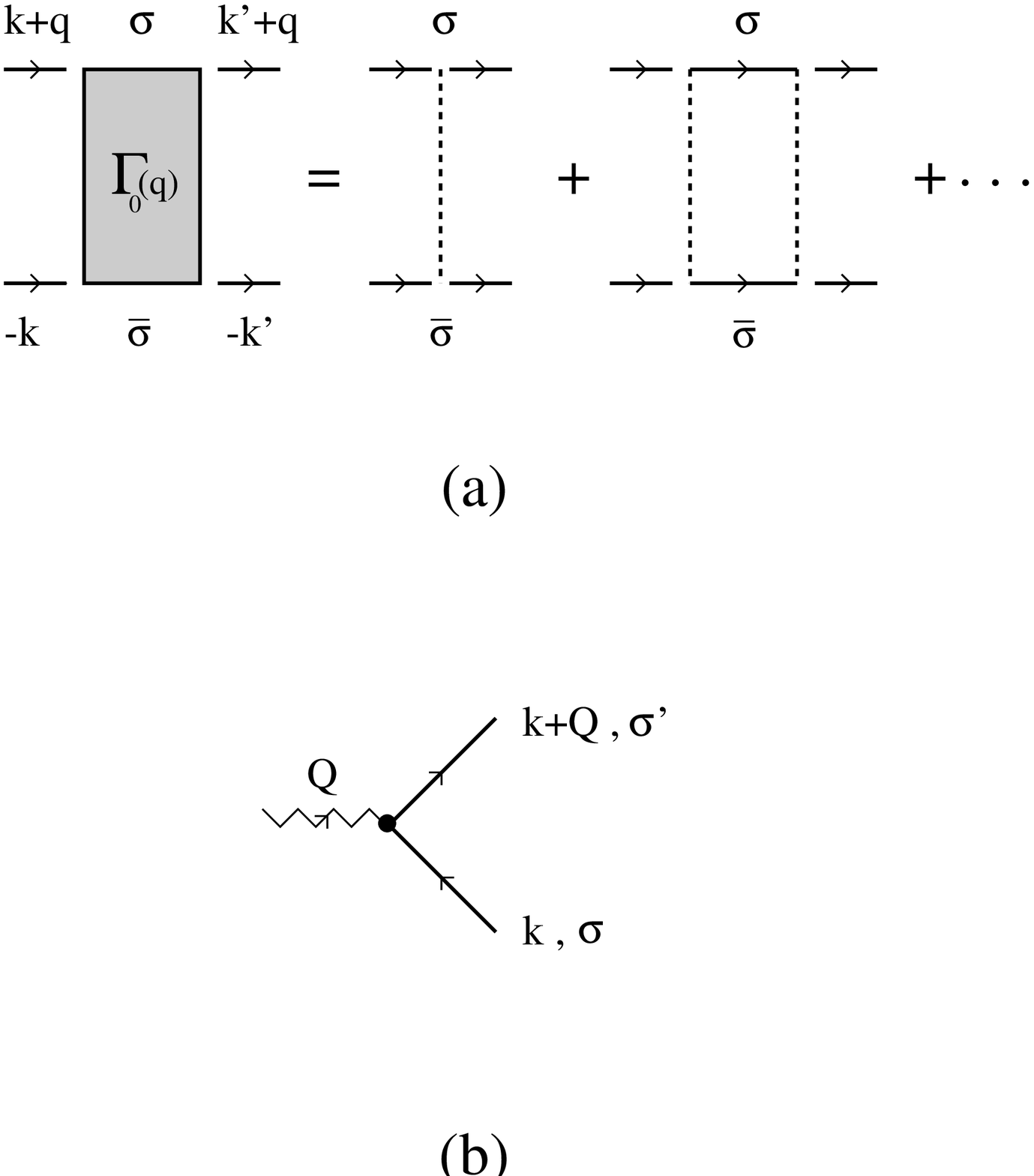}
\vspace{.4truecm}
\caption{(a) Particle-particle ladder, where full and broken lines represent fermionic
bare single-particle Green's functions and interactions, respectively (four-momenta and spin 
labels are indicated);
(b) Current (or density or spin) vertex connecting the external 
(electromagnetic) disturbance to the fermionic two-particle Green's function.}
\end{figure}

In the strong-coupling limit, $\mu$ approaches the value $- \epsilon_{0}/2$ where 
$\epsilon_{0} = (m a_{F}^{2})^{-1}$ is the binding energy of the associated two-body 
problem.
As $\epsilon_{0}$ increases without bound in strong coupling, at any finite temperature
we may consider the limit $\beta \mu \rightarrow - \infty$ in 
Eq.~(\ref{general-pp-sc}), thus obtaining the 
\emph{polar structure\/}:\cite{Pi-S-98}

\begin{equation}
\Gamma_0(q) \, \cong \, - \, \frac{8 \pi}{m^{2} a_{F}} \,\,  
\frac{1}{i\omega_{\nu} \, - \, \frac{{\mathbf q}^{2}}{4 m} \, + \, \mu_{B}}           
\label{bosonic-pp-sc}
\end{equation}

\noindent 
where $\mu_{B}= 2\mu + \epsilon_{0}$.
Apart from the residue being different from unity, this expression has the form of a 
free propagator for (composite) bosons with mass $m_{B}=2m$ and chemical potential 
$\mu_{B}$.
Note that Eq.~(\ref{bosonic-pp-sc}) holds provided
$|\omega_{\nu}| \ll \epsilon_{0}$ and ${\mathbf q}^{2}/(4m) \ll \epsilon_{0}$, which can be satisfied for all relevant values of $\omega_{\nu}$ and 
${\mathbf q}$ when $\epsilon_{0}$ is sufficiently large.

In the weak-coupling limit, on the other hand, the chemical potential is (slightly) smaller 
than the Fermi energy $\epsilon_{F}=k_{F}^{2}/(2m)$ ($k_{F}$ being the Fermi wave vector) 
for temperatures much smaller than $\epsilon_{F}$ itself.
In this case, the particle-particle ladder (\ref{general-pp-sc}) acquires the form 
characteristic of superconducting fluctuation theory:\cite{AL}

\begin{equation}
\Gamma_0(q) \, \cong \, \frac{1}{N_{0}} \,\,  
\frac{1}{\frac{T-T_{c}}{T_{c}} \, + \, \eta {\mathbf q}^{2} \, + \, \gamma |\omega_{\nu}|}
                                                                  \label{fermionic-pp-sc}
\end{equation}

\noindent
where $N_{0}$ is the free-fermion density of states at the Fermi level (per spin component), 
$(T - T_{c})\ll T_{c}$ where $T_{c}$ is the BCS critical temperature, $\gamma=\pi/(8T_{c})$, 
and

\begin{equation}
\eta \, = \, \frac{7 \, \zeta(3)}{48 \, \pi^{2}} \, 
                 \left(\frac{k_{F}}{m T_{c}}\right)^{2}                 \label{gamma}
\end{equation}

\noindent
in three dimensions ($\zeta(3) \approx  1.202$ being the Riemann zeta function of 
argument $3$).

On physical grounds, one expects the response 
functions in the strong-coupling limit of 
the original Fermi system to be expressed {\em entirely} in terms of 
composite-boson structures, namely, bosonic propagators and vertices.
As anticipated in the Introduction, the evolution of the response functions 
from 
strong to weak coupling discussed in the present paper rests on the fact that 
the particle-particle ladder (which in the strong-coupling 
limit has the form (\ref{bosonic-pp-sc}) of a composite-boson propagator) 
becomes itself the building block of fluctuation theory in the weak-coupling 
limit [cf. Eq.~(\ref{fermionic-pp-sc})].

Before identifying the relevant bosonic diagrams for the current response
function, it is useful to establish a procedure to map a given bosonic 
diagram onto a corresponding set of fermionic diagrams. To this end, 
we proceed in a {\em heuristic\/} way and formulate the following 
{\em prescriptions\/}: 
(i) Remove from the given bosonic diagram the two outer vertices 
representing the bosonic coupling to the external field, thus obtaining a 
bosonic diagram ``open'' at its ends;
(ii) Replace the bare bosonic propagators by the particle-particle ladders
given by Eq.~(\ref{general-pp-sc}); (iii) Connect the ensuing (fermionic) 
diagram to the
fermionic vertex of Fig.~1(b) representing the fermionic coupling to the external field; 
(iv) Connect eventually the remaining dangling ends of the particle-particle 
ladders among themselves, in accordance with their spin structure.

In this way, besides the fermionic diagrams which correctly reproduce the 
value of the
original bosonic diagram in the strong-coupling limit, additional fermionic 
diagrams 
may result which do not have a bosonic analogue in the strong-coupling limit 
and whose value is accordingly suppressed in this limit.
These additional diagrams will consistently be dismissed when  
mapping the original bosonic diagrams onto the associated fermionic 
diagrams.

\subsection{Leading diagrams}

For a system of noninteracting bosons, the current response function is 
depicted diagrammatically in Fig.~2(a).
This diagram represents the leading contribution to the current response 
function also for a system of bosons interacting via a (repulsive) 
finite-range potential at sufficiently \emph{low density\/}.

With the prescriptions listed above, the fermionic diagrams of Figs.~2(b) and 
2(c) are generated 
from the bosonic diagram of Fig.~2(a), with a degeneracy factor of $2$ each, 
due to the fermionic spin multiplicity.
(An additional diagram, which corresponds to a self-energy decoration of both 
bare fermionic
propagators in the fermionic particle-hole bubble, is also generated according to the
above prescriptions. Since this diagram does not have a bosonic analogue in the strong-coupling
limit, it will not be considered in the following according to the above discussion.) 
Although the two diagrams 2(b) and 2(c) are topologically not equivalent, their expressions 
coincide for particle-particle ladders corresponding to a point-contact potential.
We thus consider only one of these diagrams (say, diagram 2(b)) with a 
multiplicity factor of $4$. 

\begin{figure}
\centering
\narrowtext 
\epsfxsize=2.9in
\epsfbox{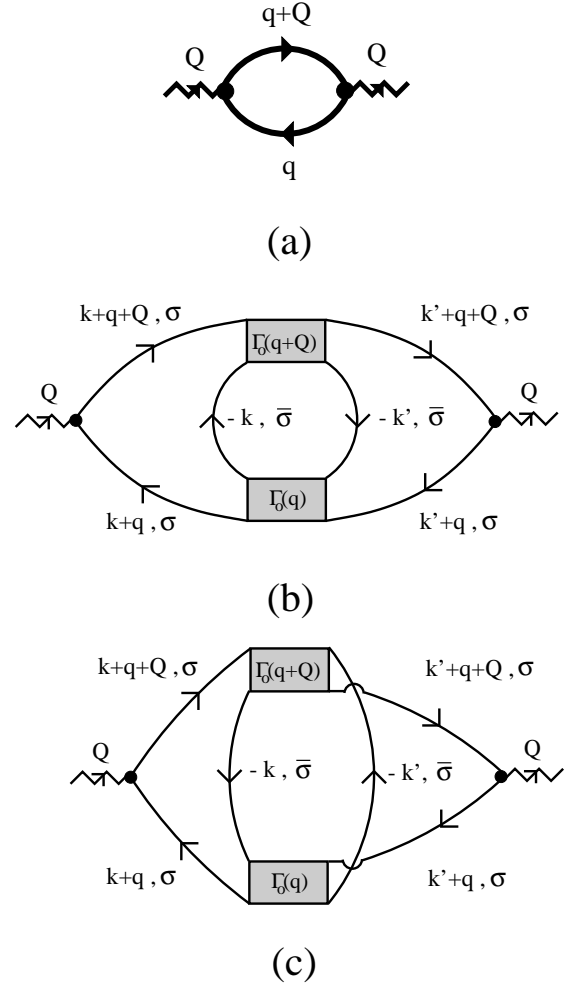}
\vspace{.4truecm}
\caption{(a) Current response function for an ideal Bose gas above the Bose-Einstein
temperature, where thick lines represent bosonic bare single-particle Green's functions;
(b)-(c) Corresponding current response function for a system of composite
 bosons.}
\end{figure}

This diagram contains two (vector) factors of the type:

\begin{eqnarray}
{\mathbf J}(q,Q)&=& \frac{1}{\beta} \sum_{\omega_{n}} \, \int \frac{d{\mathbf k}}{(2 \pi)^{3}} 
\, \frac{\left[2({\mathbf k}+{\mathbf q}) + {\mathbf Q}\right]}{2m}\nonumber\\
&\times& {\mathcal G}^{0}(-k) \,\, {\mathcal G}^{0}(k+q) \,\, {\mathcal G}^{0}(k+q+Q)    \label{bosonic-jota}
\end{eqnarray}

\noindent
where $Q\equiv ({\mathbf Q},\Omega_{\nu})$ and $k \equiv ({\mathbf k},\omega_{n})$ are bosonic
and fermionic four-vectors, respectively, and ${\mathcal G}^{o}(k)=(i\omega_{n}-\xi({\mathbf k}))^{-1}$ 
is a fermionic \emph{bare\/} propagator ($\omega_{n}=(2n+1)\pi\beta^{-1}$ ($n$ integer) 
being a fermionic Matsubara frequency).
Symmetry arguments show that ${\mathbf J}(q,Q)$ is directed along $(2{\mathbf q}+{\mathbf Q})$,
allowing us to set

\begin{equation}
{\mathbf J}(q,Q) \, = \, \frac{(2{\mathbf q}+{\mathbf Q})}{2m} \,\, C(q,Q) \,\, . 
 \label{bosonic-jota-2} 
\end{equation}

\noindent
The (scalar) factor $C(q,Q)$ can be readily evaluated in the strong-coupling limit for 
vanishing external four-vector ($Q=0$). 
In this limit, the Fermi functions originating from the sum over $\omega_{n}$ in 
Eq.~(\ref{bosonic-jota}) vanish \emph{exponentially\/} like $\exp (-\beta |\mu|)$, yielding

\begin{equation}
C(q,Q=0) \, \approx \, - \, \frac{m^{3/2}}{16 \pi} \, \frac{1}{\sqrt{2|\mu|}}
\label{V}
\end{equation}

\noindent
at the leading order in $|\omega_{\nu}/\mu|$ and ${\mathbf q}^{2}/(2m|\mu|)$.
With these approximations, and using the expression (\ref{bosonic-pp-sc}) for
the particle-particle ladder in the strong-coupling limit, 
the value of the diagram of Fig.~2(b) for $Q=0$ becomes
(the ``static'' limit with $\Omega_{\nu} = 0$ and ${\mathbf Q} \rightarrow 0$ 
is implied):

\begin{eqnarray}
\chi_{j}(Q=0) &\cong& - \, 4 \, \frac{1}{m^{2}} \, \frac{m^{3}}{(16 \pi)^{2}} 
\, \frac{1}{2|\mu|} \, \left(\frac{8 \pi}{m^{2} a_{F}}\right)^{2}\nonumber\\ 
&\times& \frac{1}{\beta} \sum_{\omega_{\nu}} \, \int  \frac{d{\mathbf q}}
{(2 \pi)^{3}} \, \frac{{\mathbf q} \, {\mathbf q}}{\left(i\omega_{\nu} \, - 
\, \frac{{\mathbf q}^{2}}{4 m} \, + \, \mu_{B}\right)^{2}}
\label{chi-sc}   
\end{eqnarray}
where the overall minus sign complies with the definition of the current response function 
$\chi_{j}$, the factor of 4 represents the degeneracy of the diagram, and the remaining factors 
stem from Eqs.~(\ref{V}) and (\ref{bosonic-pp-sc}), in the order.
Apart from the degeneracy factor of $4$, expression (\ref{chi-sc}) coincides with the $Q=0$
limit of the current response function for a system of (composite) bosons with mass $m_{B}=2m$
and chemical potential $\mu_{B}$, when for $|\mu|$ one uses the value 
$(2m a_{F}^{2})^{-1}$ which holds in the strong-coupling limit.
This response function then equals $-n_{B}/m_{B}$, where the bosonic density 
$n_{B}=n/2$ is half the original fermionic density $n$.
The degeneracy factor of $4$ in Eq.~(\ref{chi-sc}) restores eventually the correct value $-n/m$ 
for the diagonal component of the fermionic current response function, in 
accordance with the f-sum rule.\cite{Baym-1}  
This is an explicit check that the heuristic prescriptions formulated 
above lead indeed to a meaningful mapping between bosonic and fermionc 
diagrams.

It is worth noting that, when the expression (\ref{fermionic-pp-sc}) for the 
particle-particle ladder (valid in the weak-coupling limit close to $T_{c}$) 
is used in diagram 2(b) and the expression (\ref{bosonic-jota-2}) is also 
retained, one recovers the standard 
Aslamazov-Larkin contribution to the current response function,\cite{AL} which 
represents the leading fluctuation contribution in the weak-coupling limit. 
This is a nontrivial result because the weak- and strong-coupling regimes admit
entirely different classifications schemes based, respectively, on the 
Ginzburg and diluteness (gas) parameters.

\begin{figure}
\narrowtext 
\hspace{.9truecm}
\epsfxsize=2.0in
\epsfbox{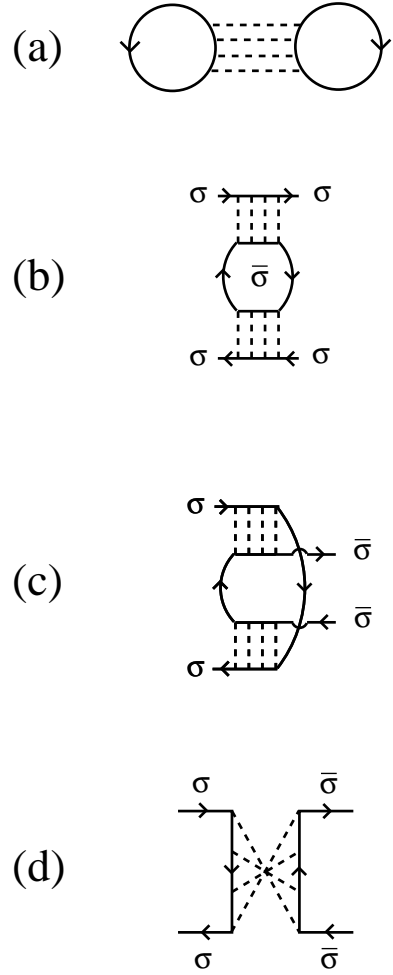}
\vspace{.4truecm}
\caption{(a) ``Potential'' from which the diagrams of Figs.~2(b) and 2(c) are
generated by the Baym's prescriptions;
(b)-(d) Fermionic effective two-particle interaction derived from the ``potential'' (a)
(with spin labels corresponding to a contact interaction).}
\end{figure}

A lowest-order scheme to interpolate between weak and strong coupling for the 
current response
function (and thus to address the interesting intermediate-coupling regime) 
can thus be set up by considering diagram 2(b) together with the fermionic 
bare particle-hole bubble.
In this way, in the weak-coupling limit one retains the free-fermion result 
plus its AL fluctuation correction, while in the strong-coupling limit one 
recovers the free-boson result,
which represent the dominant contributions in the respective limits. 

It may be mentioned that the AL diagrams 2(b) and 2(c)
can be obtained also via the Baym's prescriptions, \cite{Baym-2}
using the ``potential'' depicted in Fig.~3(a) in fermionic language.
Taking two successive functional derivatives with respect to the fermionic single-particle 
propagators (which, in this case, are meant to be self-consistent) yields, in fact, the 
fermionic effective two-particle interaction depicted in Figs.~3(b)-(d), which acts as 
the kernel of the Bethe-Salpeter equation for the two-particle Green's 
function. Diagrams 2(b) and 2(c) then result by connecting, respectively, 
diagrams 3(b) and 3(c) with the fermionic vertex of Fig.~1(b), while 
connecting diagram 2(d) with the fermionic vertex of Fig.~1(b) yields 
the Maki-Thompson diagram,\cite{MT} to be considered
in Section II.C. 

\subsection{Next-to-leading diagrams}

The leading contribution to the current response function considered in 
Section II.A corresponds to non-interacting composite bosons. The residual
interaction between composite bosons should, however, play an important role in
the not-too-extreme strong-coupling regime.~\cite{NSR,Pi-S-98}
For this reason, it is relevant to introduce the effects of the interaction
between composite bosons in the physical response functions. This
leads us to search for nontrivial corrections to the AL diagram, by examining
diagrams of higher-order in the bosonic \emph{diluteness parameter\/} 
$n_{B}^{1/3} a_{B}$, where $a_{B}$ is the scattering length associated with the residual interaction between composite bosons.\cite{footnote-1}

A bosonic diagram, which is next-to-leading with respect to diagram 2(a), is depicted in Fig.~4(a), where the dark square in the middle 
represents a (symmetrized) bosonic interaction.\cite{Popov}
The presence of an \emph{additional\/} bosonic \emph{cycle\/} in diagram 
4(a) with respect to diagram 2(a) accounts, in fact, for an additional power 
in the bosonic density.
A further bosonic diagram of the same order in the density can be obtained 
from diagram 2(a), by dressing either one of the two bosonic 
propagators with a low-density self-energy, as in the theory of the 
interacting dilute Bose system.\cite{Popov}
The physical interplay of these two diagrams in the context 
of the density and spin response functions will be addressed in Section IV. 

The bosonic diagram of Fig.~4(a) can be mapped into a corresponding set of 
diagrams for 
the fermionic response function(s), according to the rules developed in 
Ref.~\onlinecite{Pi-S-98} for the interaction vertex 
and to the heuristic prescriptions stated above.
In this way, one ends up with the two fermionic diagrams of Figs.4(b) and 
4(c), with a degeneracy factor of $8$ and $4$, in the order, having also 
taken into 
account that expressions of topologically not equivalent diagrams may coincide
for a fermionic point-contact potential.
We will verify below that, while diagram 4(b) has a meaningful 
strong-coupling limit in terms of 
composite-boson propagators, diagram 4(c) lacks a bosonic 
representation and yields consistently a subleading contribution in this limit.
For these reasons, one may retain diagram 4(b) and disregard diagram 4(c) to 
follow the evolution from strong to weak coupling.

To verify that to diagram 4(b) there corresponds a meaningful strong-coupling 
limit, we evaluate the central part of this diagram for $Q=0$ and obtain:
\begin{figure}
\narrowtext
\hspace{.3truecm}
\epsfxsize=2.9in
\epsfbox{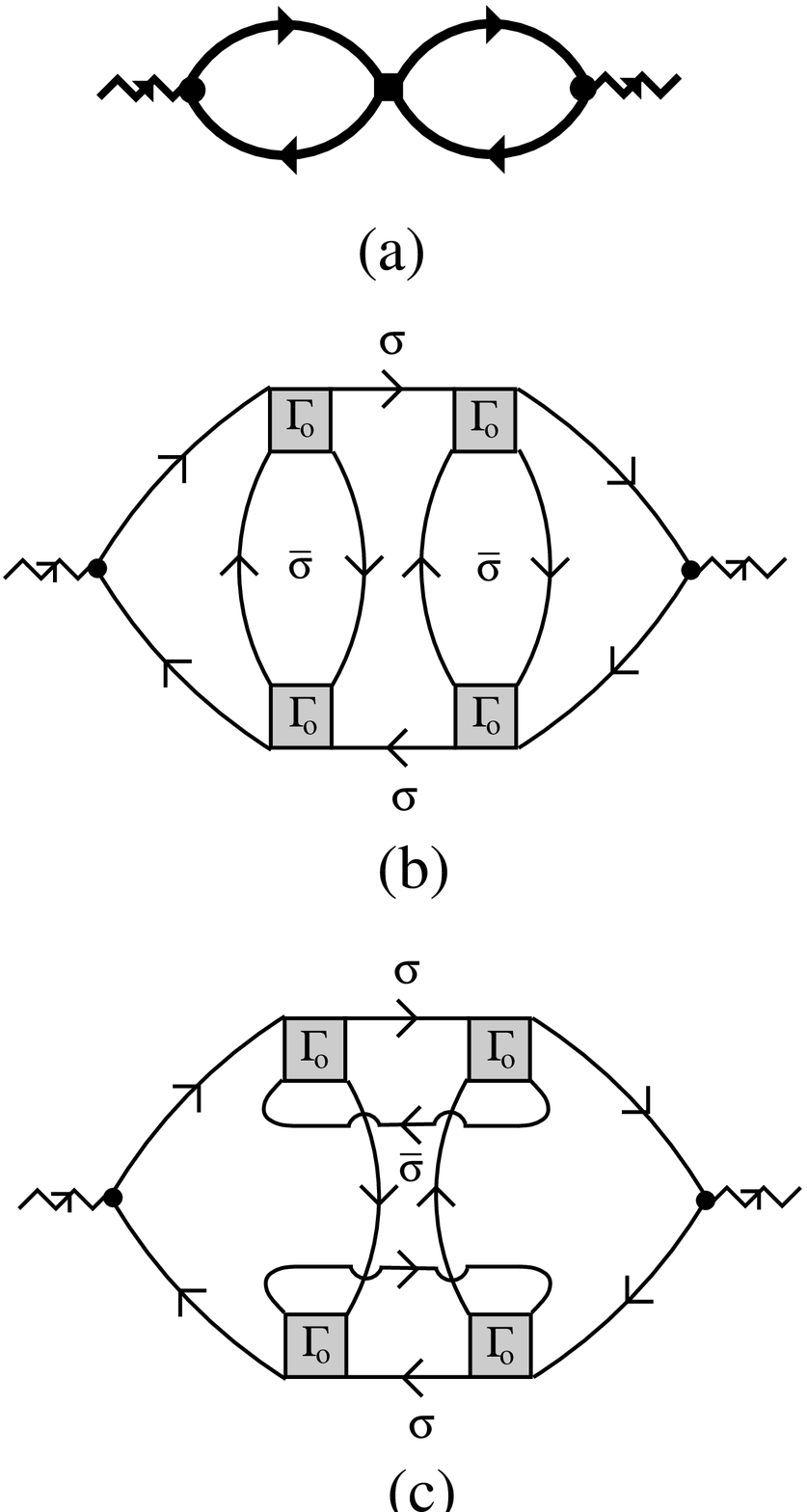}
\vspace{.4truecm}
\caption{(a) Next-to-leading bosonic diagram in the dilute limit;
(b) Fermionic diagram corresponding to the bosonic diagram (a); 
(c) Subleading diagram generated from diagram (a).}
\end{figure}
\begin{eqnarray}
& &\frac{1}{\beta} \sum_{\omega_{n''}} \, \int\frac{d{\mathbf k''}}{(2\pi)^{3}}
\,{\mathcal G}^{0}(-k'') \,\, {\mathcal G}^{0}(k''+q')\nonumber\\ 
&\times &{\mathcal G}^{0}(-k''+q-q') \,\, 
{\mathcal G}^{0}(k''+q') \, \approx \, \frac{(m a_{F})^{3}}{16 \pi}
\label{central-part}      
\end{eqnarray}
where use has been made of the relation $2|\mu| \approx \epsilon_{0} = (m a_{F}^{2})^{-1}$ that
holds in this limit.
The diagram 4(b) thus contains the factors

\begin{equation}
- \, \left(-\frac{m^{2} a_{F}}{8 \pi}\right)^{2} \, \left(-\frac{8 \pi}{m^{2} a_{F}}\right)^{4} \,
\frac{(m a_{F})^{3}}{16 \pi} \, = \, - \, \frac{4 \pi a_{F}}{m}         \label{boson-potential}
\end{equation}

\noindent
which arise, respectively, from the current vertex (cf. Eq.~(\ref{V})), from 
the residue 
of the particle-particle ladder (cf. Eq.~(\ref{bosonic-pp-sc})), and from the
 expression
(\ref{central-part}), while the overall minus sign originates from the
 three fermionic loops.
In this way, the strength $v(0) = 4 \pi a_{F}/m$ of the residual interaction 
between composite 
bosons discussed in Ref.~\onlinecite{PS} is correctly reconstructed, and
diagram 4(b) is proved to give a faithful representation of the 
bosonic diagram of Fig.~4(a).

Further, the ratio of diagram 4(c) to diagram 4(b) can be estimated to be of 
the order 
\begin{equation}
\frac{m}{a_{F}} \, \frac{\left(n a_{F}^{3}\right)^{2}}{\frac{\partial n}{\partial \mu}} \,\, ,
                                                                       \label{ratio}
\end{equation}

\noindent
which is indeed much smaller than unity in the low-density limit ($n_{B}^{1/3} a_{B} \ll 1$),
provided the compressibility $(\partial n/\partial \mu)$ does not vanish.\cite{footnote-2}

It is again interesting to mention that diagram 4(b) (with a degeneracy factor of $8$)
can alternatively be obtained in fermionic language by: Considering the fermionic effective 
two-particle interaction depicted in Figs.~3(b) and 3(c) to act \emph{twice\/} in the 
two-fermion Green's function; Connecting the ensuing four diagrams with the fermionic 
vertex of Fig.~1(b); Recognizing  the equivalence of these four diagrams; Summing 
eventually over the spin components.
Diagram 4(c), on the other hand is \emph{not reducible\/} 
in the (fermionic) two-particle channel and corresponds to a choice of the 
fermionic effective two-particle interaction different from 3(b) and 3(c). 

In this way, we have identified the next-to-leading contributions to the 
dominant (AL) diagram, which take into account correlation effects among 
composite bosons in the strong-coupling limit.
\subsection{Subleading diagrams}

We finally consider additional diagrams (besides diagram 4(c)  
considered in Section II.B)
which are subleading in the strong-coupling limit.
A noticeable example is the Maki-Thompson diagram, 
which is obtained by connecting the effective two-particle interaction of Fig.~3(d) with the 
external coupling of Fig.~1(b).
This diagram has been extensively studied in the weak-coupling limit in the context of the theory 
of superconducting fluctuations.\cite{MT,Tinkham-75,Varlamov}
Since this diagram contains \emph{only one\/} particle-particle ladder, 
it is expected to have no 
analogue in bosonic language (at least for the normal phase) and, 
consequently, not to contribute 
to the response functions in the strong-coupling limit.
Upon evaluating the MT diagram for the current response function in the 
strong-coupling
limit at $Q=0$, however, one obtains the finite value $n/m$ (including spin multiplicity).

This apparent contradiction can be overcome by considering also the 
density-of-state 
(DOS) diagram depicted in Fig.~5(a) (with a multiplicity factor of 4), obtained by
making the fermionic self-energy insertion of Fig.~5(b) into the bare 
fermionic particle-hole bubble.
In the strong-coupling limit, diagram 5(a) gives, in fact, the contribution $-n/m$ to the
$Q=0$ current response function, thus cancelling exactly the contribution of the MT 
diagram.\cite{footnote-3,footnote-4}

\begin{figure}
\narrowtext
\hspace{1.3truecm}
\epsfxsize=2.2in
\epsfbox{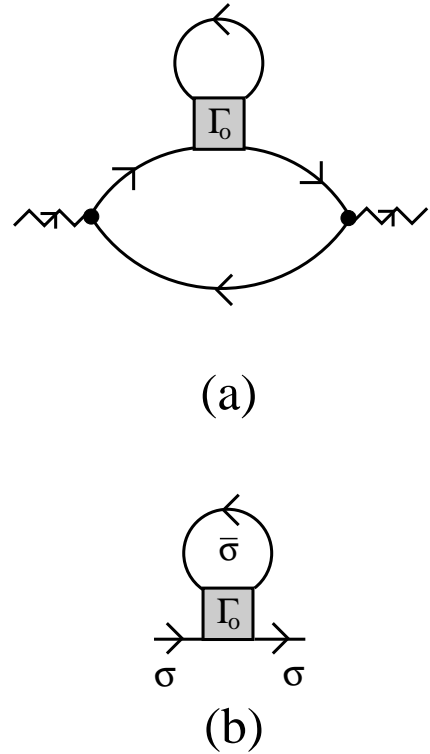}
\vspace{.4truecm}
\caption{(a) DOS diagram for the response functions;
(b) Self-energy diagram for a dilute Fermi gas.}
\end{figure}
This example suggests that diagrams for the response functions may need to 
be {\em grouped into suitable sets\/}, in order to get a meaningful 
strong-coupling limit.
The grouping procedure appears to be especially relevant for the spin response function, 
that ought to vanish in the strong-coupling limit for  
spinless composite bosons, as
discussed in the next Section.
\section{Density and spin response functions}

In this Section, we complement the analysis of the strong-coupling limit by 
analyzing the 
density and spin response functions. We begin by considering the standard
AL, MT, and DOS diagrams of the theory of superconducting 
fluctuations.~\cite{Tinkham-75,Varlamov}
We next consider a specific example to show  that a whole set of diagrams 
needs be associated with a given subleading diagram, for the spin response 
function to be 
\emph{exponentially suppressed\/}  in the strong-coupling limit, as required 
on physical grounds for spinless composite bosons.

The contribution to the density response function from the AL diagram contains two (scalar)
factors of the type

\begin{eqnarray}
D(q,Q) &=&  \frac{1}{\beta} \sum_{\omega_{n}} \, \int \frac{d{\mathbf k}}{(2 \pi)^{3}} 
\,\, {\mathcal G}^{0}(-k) \,\, {\mathcal G}^{0}(k+q)\nonumber\\
&\times&  {\mathcal G}^{0}(k+q+Q) \,\, , \label{bosonic-density}
\end{eqnarray}

\noindent
which can be readily evaluated in the strong-coupling limit for $Q=0$, to give

\begin{equation}
D(q,Q=0) \, \approx \, - \, \frac{m^{2} a_{F}}{8 \pi} \,\, .              \label{D}
\end{equation}

\noindent
This factor thus cancels the residue of the particle-particle ladder (\ref{bosonic-pp-sc}) in 
the strong-coupling limit, yielding for the density response function the 
following expression:
\begin{eqnarray}
\chi_{n}(Q) &\cong& \, - \, 4 \, \frac{1}{\beta} \sum_{\omega_{\nu}} \, \int  
\frac{d{\mathbf q}}{(2 \pi)^{3}} \,\, 
\frac{1}{i\omega_{\nu} \, - \, \frac{{\mathbf q}^{2}}{4 m} \, + \, \mu_{B}}
\nonumber\\
&\times &\frac{1}{i\omega_{\nu} + i\Omega_{\nu} \, - \, \frac{({\mathbf q} + {\mathbf Q})^{2}}{4 m} 
\, + \, \mu_{B}}  \,\, .                                                          \label{chi-n-sc}   
\end{eqnarray}

\noindent
Here, the minus sign is due to the definition of the density response function and the 
factor of 4 accounts for the degeneracy of the diagram.
In the ``static'' ($\Omega_{\nu}=0$ and ${\mathbf Q} \rightarrow 0$)  and ``dynamic''
(${\mathbf Q} = 0$ and $\Omega_{\nu} \rightarrow 0$) limits this expression 
correctly produces the values  $- 4 \partial n_{B}/\partial \mu_{B} = - 
\partial n/\partial \mu$ and $0$, in the order.

Concerning the spin correlation function, the contributions to $\chi_{zz}$ from the AL diagrams 2(b) and 2(c) cancel each other \emph{identically\/} for 
\emph{all\/} coupling strengths 
(these diagrams, on the other hand, do not contribute to $\chi_{xx}$ and $\chi_{yy}$ owing to 
their spin structure).
This is consistent with our previous result that, in the strong-coupling limit, the
AL diagram gives an appropriate description of a system of composite bosons.
One may further verify that the spin response function vanishes 
{\em identically\/} also for the corrections 4(b) to the AL diagram,  
a result which is also expected since this diagram was selected in the 
strong-coupling limit.

The contributions to the ($Q=0$) density response function from the MT 
diagram 3(d) and the DOS diagram 5(a) do not cancel each other
in the strong-coupling limit, contrary to the case of the current response function treated in Section II.C.
Rather, each of these diagrams gives the same finite contribution 
$- m a_{F}^{2} n_{B} = - n/(4 |\mu|)$, which however vanishes as $|\mu|$ 
increases in the strong-coupling limit. 
In the strong-coupling limit, both MT and DOS diagrams are thus irrelevant 
also for the density response function.

The contributions to the spin correlation function $\chi_{zz}$ from the MT and
DOS diagrams
cancel instead each other for $Q=0$ in the strong-coupling limit, since the 
MT contribution 
acquires an extra minus sign with respect to the DOS contribution.
In particular, the spin response function, obtained by considering these 
two diagrams simultaneously, vanishes \emph{exponentially\/} like 
$\exp(- \beta |\mu|)$ when approaching the strong-coupling 
limit, due to the behavior of the Fermi functions in this limit.
This is precisely what is expected on physical grounds, since a non-vanishing 
contribution to the spin 
response for spinless composite bosons results only when the temperature
is comparable with their binding energy and the composite bosons break apart.
In this context, it is interesting to mention that the progressive vanishing of the spin 
susceptibility upon approaching the strong-coupling limit has been confirmed 
by Monte Carlo 
data for the negative-U Hubbard model\cite{Andrea}, even though 
the predicted exponential behavior cannot be fully confirmed from the limited 
set of Monte Carlo data.

The above examples concerning the spin response for the AL, MT, and DOS 
diagrams (plus the correction 4(b) to the AL diagram) suggest that:
(i) Diagrams selected in the strong-coupling regime according to the
diluteness condition by considering the current response function, cannot be
used to describe the spin response function, since they would 
yield a vanishing spin response function for all couplings. 
 This implies that {\em additional} diagrams have unavoidably to be considered
for a full description of the weak-coupling regime; (ii) These additional 
diagrams introduced in the weak-coupling regime (for instance, by counting 
powers of the Ginzburg parameter as in the theory of superconducting 
fluctuations) are necessarily subleading in the strong-coupling limit, as far 
as the current and density response are concerned. However, there is {\em a 
priori}
no guarantee that they also result in an {\em exponentially} vanishing spin
response function in the strong-coupling limit, as required on physical
grounds. To make sure that this
 happens, suitable sets of diagrams need to be grouped in an appropriate 
way.~\cite{footnote-4}

As a specific example, let us consider diagram 4(c), which we concluded 
in Section II.B to be subleading as far as the current response is concerned.
This diagram alone yields a contribution to the 
spin response function which is \emph{not\/} exponentially vanishing in the 
strong-coupling limit.
Additional diagrams have thus to be associated with diagram 4(c), to obtain 
the correct exponential behavior of the spin response function in the 
strong-coupling limit.
To this end, we consider the two contributions to the thermodynamic potential depicted 
schematically in Figs.~6(a) and 6(b) and perform all possible ($Q=0$) magnetic-field insertions 
in the fermionic single-particle propagators, as to get the ``static'' spin susceptibility (no additional contributions are obtained by making 
magnetic-field insertions 
inside the particle-particle ladder in the strong-coupling limit). 
In this way, two sets of six diagrams each result, which include, by construction, diagram
4(c) (counted twice, due to the equivalence of two diagrams for the case of a 
point-contact potential) plus decorations of the AL, MT, and DOS diagrams.
In the strong-coupling limit (when all terms proportional to the Fermi 
functions are
exponentially suppressed), it can be indeed be shown that the contributions to
the spin response 
function $\chi_{zz}$ from the six diagrams obtained from Fig.~6(a) (as well from the six 
diagrams obtained from Fig.~6(b)) add up to zero.

To summarize, we have argued that diagrams which have a meaningful 
strong-coupling limit as far as the current and density response are concerned,
yield an {\em identically vanishing\/} contribution to the spin response 
function.
Other diagrams that do not have a meaningful strong-coupling limit, on the 
other hand, give contributions to the spin response function in the 
strong-coupling limit which instead vanish {\em exponentially\/} like 
$\exp(- \beta |\mu|)$ in the correct way, provided these diagrams are grouped 
into suitable sets, as shown explicitly by the examples considered above.
\begin{figure}
\narrowtext
\hspace{1.7truecm}
\epsfxsize=1.5in
\epsfbox{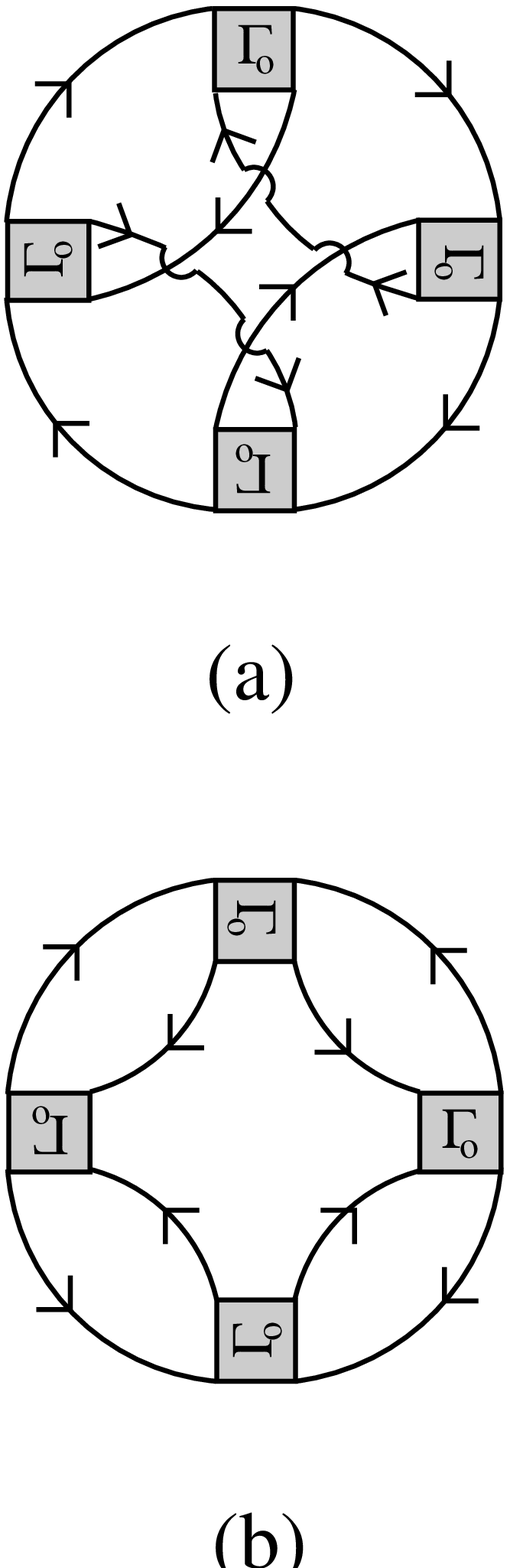}
\vspace{.4truecm}
\caption{(a)-(b) Diagrams for the thermodynamic potential, from which the 
contributions of Figs.~4(b) and 4(c) to the static spin susceptibilities can 
be derived.} 
\end{figure}

\section{Discussion and concluding Remarks}

In this paper, we have examined the evolution from weak to strong coupling of 
the response functions for a three dimensional (clean) Fermi system with an 
attractive interaction above its critical temperature.
While in the weak-coupling limit the standard analysis of superconducting 
fluctuations applies, we have shown that in the strong-coupling limit the 
original fermionic response functions become identical to the response 
functions of a system of composite bosons.
We have, in fact, verified that only those fermionic diagrams, to which there 
corresponds a meaningful representation in terms of composite bosons, 
contribute to the strong-coupling limit.
The AL, MT, and DOS diagrams of superconducting fluctuation theory have been 
analyzed among others.
We have also argued that the analysis of the spin response function may serve 
as a constraint to select sets of diagrams for the current and density response
functions, which are relevant for weak coupling but are suppressed for 
strong coupling.

It is evident from our analysis that many diagrams contributing 
to the weak-coupling limit are suppressed in the strong-coupling limit.
Consistently, by selecting the relevant diagrams for the response functions 
starting only from the strong-coupling limit, one might miss important 
contributions to the weak-coupling limit.
For this reason, our analysis in the strong-coupling limit must be supplemented
by the standard criterion of superconducting fluctuation theory for selecting
suitable sets of diagrams in the weak-coupling limit.
This is especially true for the spin response function, which vanishes for a 
system of spinless bosons: Extrapolating to the weak-coupling limit only 
diagrams which contribute in the bosonic limit to the current and density 
response functions, would result into a vanishing spin response function 
for \emph{all\/} coupling strengths.

Controlling the two (weak- and strong-coupling) limits separately may prove 
especially important for describing the intermediate (crossover) region, for 
which no controlled theory can be specifically formulated.
One reasonable strategy to approach the crossover region is then to 
\emph{interpolate\/} between theories which are  controlled in the two (weak- 
and strong-coupling) limits,
which can be done by including all dominant diagrams in either one 
of the two limits and then evaluating them over the whole coupling 
range. This contrasts somewhat with what was found in 
Ref.~\onlinecite{Pi-S-98} for the fermionic self-energy, for which a
single approximation selected in the strong-coupling regime proved also 
sufficient to describe the weak-coupling region.
For the response functions, at the leading order one may include the AL 
diagram (which is dominant both in the strong- and weak-coupling limit) plus 
the MT and DOS diagrams (which are relevant to the weak-coupling limit but are 
strongly suppressed in the strong-coupling limit). At the 
next-to-leading order, the effect of the residual interaction between 
composite bosons can be included considering the corrections to the AL 
diagrams discussed in Section II.B.

In this context, it is interesting to comment on the recent results reported 
in Ref.~\onlinecite{Perali} regarding the temperature dependence of the 
density and spin susceptibilities for a two-dimensional negative-U Hubbard 
model, calculated via the AL, MT and DOS diagrams, and then compared
with available Monte Carlo results for $U=-4t$ ($t$ being the nearest-neighbor
hopping).
These authors find a remarkable agreement between their calculation and the 
Monte Carlo data for the spin susceptibility, {\em provided\/} the mass term 
in the particle-particle ladder (\ref{fermionic-pp-sc}) is replaced by a mass 
term with the characteristic temperature dependence of the Kosterlitz-Thouless
theory (this replacement should amount to inserting self-energy corrections
in the bosonic propagators of the AL diagram).
For the density susceptibility, however, this replacement alone proved not 
sufficient to reproduce the Monte Carlo data.
The discussion presented in Section II.B indeed suggests that modifications 
of the AL diagram obtained by considering bosonic self-energy corrections to 
the particle-particle ladder should also be accompanied by the inclusion of 
an additional diagram (namely, diagram 4(b) for the density response function),
which in the strong-coupling limit accounts for the residual bosonic 
interaction at the \emph{same\/} order in the 
diluteness parameter.
Numerical calculations including this additional diagram have not 
yet been performed.

In this paper, we have considered the response functions in the normal phase 
\emph{above\/} the critical temperature.
It would certainly be interesting to extend this analysis \emph{below\/} the 
superconducting
critical temperature and study the continuous evolution of the response functions from the 
weak-coupling limit of (BCS) superconductivity to the strong-coupling limit where Bose-Einstein 
condensation takes place.
In this case, a description in terms of Bogoliubov quasi-particles may be
appropriate for a dilute system of composite bosons (at least close to zero temperature), 
with the superfluid density being affected at finite temperature by sound modes in the
strong-coupling limit and by pair-breaking effects in the weak-coupling limit.
Which of these two effects dominate in the intermediate (crossover) region is a challenging 
question, which can be addressed only by numerical calculations of a suitable set of diagrams.
Work along these lines is in progress.

\acknowledgments

The authors are indebted to C. Castellani, A. Perali, and A. Varlamov for 
helpful discussions.
Partial financial support from the Italian INFM under Contract PAIS Crossover 
No.~269 is gratefully acknowledged.



\begin{references}

\bibitem{Schrieffer} J.R. Schrieffer, \emph{Theory of Superconductivity\/}
                     (Benjamin, New York, 1964).

\bibitem{FW} See, e.g., A.L. Fetter and J.D. Walecka, \emph{Quantum Theory of
 Many-Particle Systems\/} (McGraw-Hill, New York, 1971), Chapter 13.

\bibitem{AL} L.G. Aslamazov and A.I. Larkin, Sov. Phys. JETP {\bf 10}, 875 
(1968).

\bibitem{MT} R.S. Thompson, Phys. Rev. B {\bf 1}, 327 (1970);
             J.P. Hurault and K. Maki, Phys. Rev. B {\bf 2}, 2560 (1970).

\bibitem{Tinkham-75} W. J. Skocpol and M. Tinkham, Rep. Prog. Phys. {\bf 38}, 
1049 (1975).

\bibitem{Varlamov} A.A. Varlamov, G. Balestrino, E. Milani, and D.V. Livanov,
Adv. Phys. {\bf 48}, 655 (1999).
              
\bibitem{Ding} H. Ding \emph{et al.\/}, Nature {\bf 382}, 51 (1996); and
               Phys. Rev. Lett. {\bf 78}, 2628 (1997).

\bibitem{Loeser} A. G. Loeser \emph{et al.\/}, Science {\bf 273}, 325 (1996).
                  
\bibitem{Leggett} A.J. Leggett, in \emph{Modern Trends in the Theory of 
Condensed Matter\/}, edited by  A. Pekalski and R. Przystawa, Lecture Notes in
Physics Vol.115 (Springer-Verlag, Berlin, 1980), p.13. 
                  
\bibitem{NSR} P. Nozi\`{e}res and S. Schmitt-Rink, J. Low. Temp. Phys.
 {\bf 59}, 195 (1985).                                   

\bibitem{Randeria} C.A.R. S\'a de Melo, M. Randeria, and J.R. Engelbrecht,
                      Phys. Rev. Lett. {\bf 71}, 3202 (1993).

\bibitem{Haussmann} R. Haussmann, Z. Phys. B {\bf 91}, 291 (1993).

\bibitem{PS} F. Pistolesi and G.C. Strinati, Phys. Rev. B {\bf 49}, 6356 
(1994); \emph{ibid.\/} B {\bf 53}, 15168 (1996).
             
\bibitem{Zwerger} S. Stintzing and W. Zwerger, Phys. Rev. B {\bf 56}, 9004 
(1997).

\bibitem{Levin} B. Jank\'o, J. Maly, and K. Levin, Phys. Rev. B {\bf 56}, 
R11407 (1997).
               
                
\bibitem{Pi-S-98} P. Pieri and G.C. Strinati, Phys. Rev. B {\bf 61}, 15370 
(2000), and cond-mat/9811166.

\bibitem{Perali} L. Benfatto, A. Perali, C. Castellani, and M. Grilli, Eur. 
Phys. J. B {\bf 13}, 609 (2000).                   

\bibitem{Popov} V.N. Popov, \emph{Functional Integrals and Collective 
Excitations\/} (Cambridge University Press, Cambridge, 1987).                  
                  
\bibitem{Baym-1} G. Baym, in \emph{Mathematical Methods in Solid State and 
Superfluid Theory\/}, R.C. Clark and G.H. Derrick, Eds. (Oliver and Boyd, 
Edinburg, 1967).
                 
\bibitem{Baym-2} G. Baym, Phy. Rev. {\bf 127}, 1391 (1962).                 

\bibitem{footnote-1} It was shown in Ref.~\onlinecite{Pi-S-98} that the 
scattering length $a_{B}$ for composite bosons is proportional to the 
fermionic scattering length $a_{F}$, with a coefficient of order unity.
               
\bibitem{footnote-2} In the strong-coupling (bosonic) limit, one may write 
$\partial n/\partial \mu = 4 \, \partial n_{B}/\partial \mu_{B}$ and
estimate $\partial n_{B}/\partial \mu_{B}$ from the relation $\mu_{B}(n_{B})$
valid for an ideal Bose gas close to the Bose-Einstein temperature $T_{BE}$ 
(cf. Ref.~\onlinecite{FW}, Section 5). One obtains $\partial \mu_{B} 
/\partial n_{B} \approx 1.63 (T-T_{BE})/n_{B}$, yielding
a divergent $\partial n_{B}/\partial \mu_{B}$ as $T$ approaches $T_{BE}$,
which further suppresses diagram 4(c) with respect to diagram 4(b).
                     
\bibitem{footnote-3} In the strong-coupling (bosonic) limit, the $Q$ 
dependence of the MT and DOS diagrams are expected to be irrelevant insofar as
the fermionic chemical potential constitutes the largest energy scale in the 
problem.
                     
\bibitem{footnote-4} A similar cancellation has been noted in the 
weak-coupling limit by D.V. Livanov, G. Savona, and A.A. Varlamov, Phys. Rev. 
B (in press).                    
                     
\bibitem{Andrea} M. Randeria, N. Trivedi, A. Moreo, and R.T. Scalettar, Phys. Rev. Lett. {\bf 69}, 2001 (1992); J.M. Singer \emph{et al.\/}, Phys. Rev. B 
{\bf 54}, 1286 (1996).             
\end{references}
\end{document}